# Biological Computation as the Revolution of Complex Engineered Systems

Nelson Alfonso Gómez Cruz and Carlos Eduardo Maldonado

*Modeling and Simulation Laboratory*
*Universidad del Rosario*
*Calle 14 No. 6-25, Bogotá, Colombia*

**ABSTRACT:** Provided that there is no theoretical frame for complex engineered systems (CES) as yet, this paper claims that bio-inspired engineering can help provide such a frame. Within CES bio-inspired systems play a key role. The disclosure from bio-inspired systems and biological computation has not been sufficiently worked out, however. Biological computation is to be taken as the processing of information by living systems that is carried out in polynomial time, i.e., efficiently; such processing however is grasped by current science and research as an intractable problem (for instance, the protein folding problem). A remark is needed here: P versus NP problems should be well defined and delimited but biological computation problems are not. The shift from conventional engineering to bio-inspired engineering needs bring the subject (or problem) of computability to a new level. Within the frame of computation, so far, the prevailing paradigm is still the Turing-Church thesis. In other words, conventional engineering is still ruled by the Church-Turing thesis (CTt). However, CES is ruled by CTt, too. Contrarily to the above, we shall argue here that biological computation demands a more careful thinking that leads us towards hypercomputation. Bio-inspired engineering and CES thereafter, must turn its regard toward biological computation. Thus, biological computation can and should be taken as the ground for engineering complex non-linear systems. Biological systems do compute in terms of hypercomputation, indeed. If so, then the focus is not algorithmic or computational complexity but computation-beyond-the-Church-Turing-barrier. Computation at the biological scale is carried out as the emergence of properties, attributes, or systems throughout chemical processes wherein the usual serial, parallel, and distributed processing occupy a less relevant importance than expected. We claim that we need a new computational theory that encompasses biological processes wherein the Turing-Church thesis is but a particular case.

## 1. INTRODUCTION

We need a theoretical frame for complex engineered systems (CES). Even though in the past there have been disciplines with just medium-range theories and engineering has borrowed its theoretical presuppositions from physics and mathematics, mainly, we can reasonably think that once the shift is produced from classical and conventional engineering towards CES the situation can and may change.

The core problem concerning CES is related to the processing of information, since the basis of CES is not matter and energy any longer as it is indeed the case for conventional and classical engineering. Rather, CES is concerned with information and computational problems. If so, biological computation emerges as the ground linking bio-inspired engineering together with CES for what is really at stake is the understanding of non-linear phenomena, behaviors and dynamics.

In computational terms, the Church-Turing thesis (CTt) is the prevailing paradigm. P versus NP problems can be viewed as the result of computation as a Turing machine and the possibilities and constraint that poses a Turing machine. However, once the interest turns to biological systems the validity or extension of the CTt becomes limited. Hypercomputation emerges as both the basis and framework for the understanding of biological systems. We claim that biological systems compute in terms of hypercomputation.

We shall argue in favor of a road to CES through four arguments, thus: first we discuss the standard interpretation of the Church-Turing thesis and move towards the contributions of Turing himself for understanding life and computation. Secondly, a short study of the P versus NP problems is necessary in order to clear up the road. Consequently we claim that life does not compute deterministically and that´s why we need to move to a more fundamental level, namely hypercomputation. This is precisely the third argument. In biological systems the processing of information happens in a radically different way as in conventional computation. Thereafter, as a final argument we encounter a complexification of computation that allows us to better grasp what living systems do vis-à-vis information. The road, we believe, to CES is widely open. At the end, a few concluding remarks are formulated.

## 2. THE CHURCH-TURING THESIS, REVISITED

The Church-Turing thesis (CTt) arises in the midst of a physical paradigm. Thereafter its claim about determinism, top-down computation and the von Neumann architecture become necessary references herein. However, it should be mentioned that Turing himself never said the Turing machine was the model that could express all possible computations. That saying has been attributed to Turing and is truly a sort of uncritical claim in the community of scientists and academicians.

Indeed, together with the traditional Turing machine and consequently the importance of the CTt, others forms of Turing machines must be recognized and worked out, namely

**CORRESPONDING AUTHOR:** C. E. Maldonado, Ph. D, Full Professor, Universidad del Rosario, Bogota, Colombia; Research field: complex systems, non-linear sciences, non-classical logics, complexity theory; E-mail: carlos.maldonado@urosario.edu.co.

choice-Turing machines (c-machines), oracle-Turing machines (o-machines), and unorganized-Turing machines (u-machines). These three kinds of Turing machines succeed in overcoming the computational power of the universal Turing machine throughout the interaction of c-machines and o-machines, and via evolution as it is the case with u-machines [1]. As a consequence, the distinction must be made between the strong interpretation of the Turing machine and the very thesis itself.

Radical as it is, some argue that the Turing machine theory is not so much wrong as irrelevant [2]. The possibility of the relevance of the theory entails a revision of the theory which is not just possible but compulsory. Such a revision however goes along the way opened by Turing himself.

The CTt is the clear and straightforward claim that computing is a relevant physical process. Therefore, the distinction is made between the entering program, the very process of computation which happens as it is in and as a black closed box, and the output which is the translation of the binary Boolean program that enters early on. Living systems do not process information in this way, at all.

Indeed, for a living system processing information is one and the same process and dynamics as living. And in processing information her life is ultimately at stake. Were the processing of information a black closed box would then sooner be the living system at risk if not peril.

Engineering the processing of information of living systems is a matter of pride and prestige from a scientific and academic point of view, but also a matter of survival from an evolutionary standpoint. For if we cannot understand what living systems do for living a little favor will do scientists and engineers to the ongoing bottle-neck situation of nature and mankind. CES is ruled by the CTt, though.

In order to fully understand the processing of information of living systems a twofold way opens up, thus: either both the possibilities and constraints of the c-machines, the o-machines and the u-machines, or shifting the conventional universal (i.e. 'general purpose machine') Turing machine. The more efficient road taken by the community of scholars, researchers and theoreticians was clear enough: hypercomputation [3]. It should be pointed out however that hypercomputation remains a theoretical or conceptual suggestion, for no engineering process has been implemented in terms of hypercomputation as yet. The core reason for the understanding of the information processing is: living systems are learning systems, and thereafter adaptive and evolving systems. If so, the question becomes not just improving the conventional universal deterministic Turing machine (via, f.i., the o- u, and c-machines) but transforming the very Turing machine. That is exactly what hypercomputation is about [4].

## 3. A BRIEF CONSIDERATION: P AND NP PROBLEMS

While the theory of computation deals with distinguishing computable problems from non-computable ones within the CTt context, the (computational) theory of complexity is occupied, in the same framework, with determining the necessary resources to achieve the computation it is concerned with. This means that the theory of complexity is concerned with computable problems and how efficiently they can be solved. Thus, computability and complexity are to be seen as the two hardcore problems of the theory of computation.

Now, within the theory of complexity two central problems are to be differentiated, namely those that can be solved with accuracy in a polynomial time throughout a deterministic Turing machine (P problems a.s.a. easy problems) as well as those that can be solved in an approximate way, in a polynomial time too, throughout a non-deterministic Turing machine (NP problems, or difficult problems) [5]. The problems that require a non-polynomial time in order to be solved exactly are known as intractable problems [6].

Provided that life can be calculated algorithmically (i.e. via a Turing machine), it should become clear that life should not be understood in terms of P problems. Is life, therefore, an NP problem? The answer to this question –one more time: in terms of limits of a Turing machine, is a rotund: yes [7]. For, indeed solving the problem concerning life is something that cannot be carried out in precise or accurate terms. Besides, when looking at the models for solving biologically computational motivated problems (we refer here to the work on metaheuristics biologically motivated [5]. The most conspicuous cases are evolutionary computation, swarm intelligence, artificial immune systems, and membrane computing) life does not solve problems in a deterministic way; life does not, in any case, have a unique answer but a space of solutions.

However, such a description of life is incomplete not to mention incorrect. For, on the one hand the way biological systems process information appears to resist an explanation in terms of algorithms and, on the other hand, the problems that concern life are not clearly defined and/or delimited as it is indeed the case for NP problems. Understanding life entails, thereafter, carrying on computation at the next level, namely hypercomputation. This is exactly the kind of discussions that lie at the center of biological computation.

## 4. THE CORE: BIOLOGICAL COMPUTATION

Biological computation refers to the fact that biological systems carry out computations on their own in the midst of their natural environment [8]. In other words, biological computation has to do with processing information that is naturally carried out by living organisms. This idea is closely related to the fact that in general complex systems carry out computations [9]. Examples run from biochemical to bio-molecular systems such as DNA and RNA passing through the gene assembly in single-celled organisms or the capacity of the immune system to adapt and respond in mammals on to the dynamics of development and evolution. Therefore, it is not a sheer and useful metaphor extracted from biology that is used here as a source of inspiration in order to solve computationally complex systems in the frame of the sciences of computation (which is, for instance, the case for bio-inspired computation). It is neither the case about using computers to store, manipulate, integrate or analyze experimental biological data (which is rather the interest that have bioinformatics and the mainstream of computational systems biology). Rather, it is here the case of understanding computation as a truly fundamental phenomenon –beyond the very sense of the universal Turing machine that crosses our current computing, physical, and biological systems including social systems.

In biological systems, the processing of information happens in a radically different way as it is conceived by the current theory of computation [10]. Whereas digital computers based on the Turing machine and on the Von Neumann architecture make use of complex processors capable of carrying out complicated tasks in a sequential form [11], computation in biological systems emerges in a coordinate and de-centralized interaction of elements that are relatively simple such as, f.i., molecules, cells, organisms [12, 13]. This means that information that comes out from computation is not stored in any particular system but it manifests in the global configuration of the system. In the same tenure, the mechanisms by which computation is carried out are neither to be found nor represented in an explicit form but, instead, emerge from non-linear interactions of the elements that make up the system.

The large majority of computational models and architectures that are biologically motivated fall within the frame of the strong Church-Turing thesis and match - not without a big merit, to leveling the computational power of the Turing machine. Examples can be found in membrane computation [14], the assembling of genes [15], cellular computation [13], cellular automata [16], and DNA computation [17]. The works that from a theoretical standpoint seek to solve problems that are currently considered as non-computable are much less common – where usually the referent of the non-computable problem is the halting Turing problem. Some of these proposals are based on cellular automata, Turing evolving machines [18], analog neural networks [19] as well as on accelerated P systems [20].

The latter has meaningful implications for theoretical computation and yet it is not successful in expressing clearly the problem concerning biological computation. On the one hand, because processes such as the immune defense system, development or adaptation are clearly non-algorithmic; considering the opposite would be as if we thought that the Turing machine is capable of computing anything that is computable, including even life. On the other hand, because biological systems process information in real time and for living systems the problem concerning the Turing halting problem ceases to be a central subject [21]. Indeed, living systems do not merely exist in time (f.i. circadian cycles), but they set out time by themselves. Accordingly, biological computation is necessarily and unavoidably motivating a re-definition of computation and computability.

Every known biological system processes information on the molecular scale. Biochemical processing of information is characterized by being tolerant to failure, it is robust, resilient, self-organized, adaptive, a-synchronic, de-centralized, and evolving [22]. All these features, desirable in and characteristic of CES are far from being rightly understood, formalized and harnessed from the computational point of view. CES, we argue, depends directly from the complexification of computation, whose most salient feature is in the foreseeable future biological computation.

The advances and interests nonetheless, there is no consensus among the scientific community concerning whether biological systems process information or not. There are some researchers who in the context of computational intelligence, soft computation or the design of metaheuristics take as inspiration biology as a source for solving engineering-like problems or use computational tools on a biological plane without realizing whether the underlying phenomenon is essentially computational or not (see, for instance, [23, 24]). Others simply do not accept the fact that nature, and particularly living systems, process information, except as a metaphor [25, 26]. Finally, there are some others who have begun to assess openly that nature, and life in particular, do indeed process information [8, 15, 22, 27-30]. There are some who have even come to claim that computation is an essential (= fundamental) feature for life to be [9, 22, 31].

Biological computation arises in the frontier of spearhead science and research in the world and yet it is far from being fully understood and accepted [26]. In fact, the main efforts that link biologists and computing scientists have been oriented so far in different directions to those from biological computation. Understanding how life processes information, i.e. how it computes will allow us to understand biology in more unified way [8] while widening up our conception about what it really means "computation" [15]. Such an understanding will also have a great impact in the way how engineering and a number of disciplines work and think.

## 5. THE COMPLEXIFICATION OF COMPUTING AND COMPLEX ENGINEERED SYSTEMS

The Turing machine does not describe all possible computations but only those that are effectively computable – that is, those that are computable via mechanic or algorithmic means. More exactly, a Turing machine is related to an automatic solution of mathematical computable functions. Algorithmic computation is by definition closed in the sense that nothing enters or comes out during computation [19], [32]. Hence, the evolution of computation has consisted so far, in general terms, in storing and processing in an increasingly better way algorithmic procedures. The miniaturization of devices, the emergence of new paradigms in programming, fields such as parallelism or massive parallelism and the contributions of classic artificial intelligence to the sciences of computation point exactly in this direction.

However, how and when does the processing of information take place is something that cannot be restricted to the plane of mathematical computable functions. And it cannot be limited, not even, to the mathematics and logics in general and certainly not in the classical sense. It becomes then evident that the definition of what is, and is not, computable depends exactly on the model of computation chosen and, by extension, that there is no a unique model of computation. Moreover, not every model of computation must be necessarily equivalent to a universal Turing machine.

New models of computation haven been set out or suggested that extend the power of algorithmic computation – a conspicuous case being interactive computation [32]. In any case, all the best contributions to new models come from physics and biology. Natural computation and artificial life have contributed enormously to the shifting of the traditional understanding of what it means computing.

In contrast to artificial intelligence, artificial life did not focused on the prevailing computational paradigm but it set out the need for formulating and developing a new paradigm within computation more accordingly to its natural biological counterpart [33]. The rationale of artificial life was and has remained a theoretical one, namely understanding and explaining life, not just as it is, but also as it-could-be. Such a theoretical enterprise has modified or led to a shift in the very way how computation is understood and worked out. Artificial life has contributed significantly to the understanding of life thanks to computation [33-36]. Natural computation shares the merit of contribute to the understanding of life along the way opened up by artificial life, too [15, 29].

Not without reason, the search and construction of a computational theory of biological systems became one of the key open most important problems in the field of artificial life [37]. Working throughout synthesis, i.e. synthesizing, building from bottom up, decentralized processing and vastly parallel processing of information and ideas about emergence and self-organization are but some of the fundamental contributions that artificial life has made onto the sciences of complexity and CES. Moreover, many of the advances so far in unconventional computation as wells as in bio-inspired engineering and biological computation have been motivated by the very philosophy of artificial life.

The complexification of computation comes then from two different directions that are just beginning to converge: one is hypercomputation, and the other is biological computation. The idea, we claim, which is a scandal for tradition, is that computing beyond the strong CTt matches perfectly with the idea according to which computing life goes far beyond a sheer computationalist view.

Surely the most important contribution of artificial life (and of a number of other closely related fields such as natural computation, cellular computation or synthetic biology) to the theoretical framework of CES is not so much on the plane of metaheuristics and bio-inspired computation as, more deeply and solidly, in the works with biological computation.

The core of CES cannot be unlike classical and conventional engineering the simple manipulation and transformation of matter and of the sources of energy to benefit mankind. Even though they still play an important role in engineering and science as yet, the axis is moved toward information and information processing [38]. As of today, we argue understanding and explaining the complexity of engineering-like systems will depend directly and proportionately on the very complexification of computation. As it has been said, such a complexification is grounded on hypercomputation and biological computation.

## 6. CONCLUDING REMARKS

The most important scientific, philosophical and cultural meaning of CES has to do with both understanding and engineering complex systems. As it is well known, the best way to understand any problem consists in building it. Therefore, designing, engineering and producing become more relevant than ever when facing complex systems. To be sure, the most complex systems ever on earth are living systems.

According to what precedes, it is reasonable to conceive that both understanding and engineering a system according to how it processes information ought to be taken as a new frame of computational sciences. If so, on the one hand such a system can be recognized as complex and, on the other hand, engineering can be thought of as a complex science. In any case, living systems can be easily recognized as the ones that best and most process information. This is the reason why we argue here that biological computation is to be taken as the road to CES in general and, particularly, as the theoretical framework for CES.

The information processing has three moments throughout history: i) Shannon´s approach which consists in measuring the degree of noise a message or information carries out or is crossed by (noise, i.e. black, white and pink noise). This leads to Shannon´s entropy; ii) According to Zurek, the core of a material system consists in the information carried out and processed by the system: it comes from bit; iii) finally, thanks to Wheeler it is the quantum processing what becomes truly meaningful: bit comes from qu-bit. In this general frame, we think that living systems perceive, store, process, carry out, recognize or forget information in quantum terms. The move from information processing to computation implies a radical shifting of the latter in the sense of the transition from the CTt to hypercomputation. With this paper we aimed at paving the road toward hypercomputation. The gate seems to us the study, understanding, and engineering of living systems. The thread is clearly what distinguishes living beings from non-living systems, namely the information processing.

Computation for living beings is a question of life or death. Therefore, computational sciences together with engineering must enrich or enlarge their rods and concepts. Biological computation, hypercomputation, complex computations, biochemical processes, and some non-classical logics emerge as new tools and explications for what a living being is and can do to adapt and survive in nature. We encounter here new horizons for research and work.